\documentclass{emulateapj}
\usepackage{apjfonts}
\usepackage{bm}
\usepackage{amsmath}
\usepackage{rotating}

\newcommand\lsim{\mathrel{\rlap{\lower4pt\hbox{\hskip1pt$\sim$}}
        \raise1pt\hbox{$<$}}}
\newcommand\gsim{\mathrel{\rlap{\lower4pt\hbox{\hskip1pt$\sim$}}
        \raise1pt\hbox{$>$}}}

\shorttitle{Hot Accretion With Saturated Conduction}
\shortauthors{Menou}

\begin{document}

\title{Hot Accretion With Saturated Conduction}

\author{Kristen Menou}
\affil{Department of Astronomy, Columbia University, 550 West
120th Street, New York, NY 10027}

\hspace{\baselineskip} %\submitted{To be submitted to \apjl}

\begin{abstract}
Observations of the hot gas surrounding Sgr A* and a few other nearby
galactic nuclei imply electron and proton mean free paths comparable
to the gas capture radius: hot accretion likely proceeds under
weakly-collisional conditions in these systems. As a result, thermal
conduction, rather than convection, may be important on all scales and
affect the global flow properties. The self-similar ADAF solution of
Narayan \& Yi (1994) is generalized to include a saturated form of
thermal conduction, as is appropriate for the weakly-collisional
regime of interest. Conduction provides extra heating and yet it
reduces the free-free radiative efficiency of the accretion flow (by
potentially large factors). These idealized solutions suggest that
thermal conduction may be an important physical ingredient to
understand hot accretion onto dim accreting black holes. Conduction
could also play a role in reducing the rate at which black holes
capture ambient gas and in providing an evaporation mechanism for an
underlying cold thin disk.
\end{abstract}

\keywords{accretion, accretion disks -- conduction -- black hole
physics -- hydrodynamics}

\section{Introduction}

Over the past decade, X-ray observations have made it increasingly
clear that black holes are capable of accreting gas under a variety of
distinct configurations. There is convincing evidence for a hot form
of accretion occurring at sub-Eddington rates in particular, which
contrasts with the classical ``cold and thin'' accretion disk scenario
(Shakura \& Sunyaev 1973). Hot accretion appears common in the
population of supermassive black holes in galactic nuclei and during
the quiescent phases of accretion onto stellar-mass black holes in
X-ray transients (e.g. Narayan et al. 1998; Lasota et al. 1996; Di
Matteo et al. 2000; Esin et al. 1997, 2001; Menou et al. 1999; see
Narayan, Mahadevan \& Quataert, 1998; Narayan 2003 for reviews).

Even today, however, the exact nature, structure and properties of
these hot accretion flows remain controversial.  Inspired by the work
of Shapiro, Lightman \& Eardley (1976), Narayan \& Yi (1994; 1995a,b)
derived self-similar ADAF solutions which emphasized the important
stabilizing role of radial heat advection (see also Ichimaru 1977;
Rees et al. 1982; Abramowicz et al. 1988).  Subsequent analytical work
on hot accretion flows has emphasized outflows, motivated by a
positive Bernoulli constant (ADIOS; Blandford \& Begelman 1999), and
the potential role of convection (CDAFs; Narayan et al. 2000, Quataert
\& Gruzinov 2000). Hydrodynamical and MHD numerical simulations have
also greatly contributed to the subject by highlighting important
dynamical aspects of the problem (Hawley et al. 2001; DeVilliers et
al. 2003; Igumenshchev et al. 2003).

The purpose of this {\it Letter} is to make the case that thermal
conduction, which has been a largely neglected ingredient, could
affect the global properties of hot accretion flows substantially. In
\S2, we use existing observational constraints on a few nearby
galactic nuclei to argue that hot accretion is likely to proceed under
weakly-collisional conditions in these systems, thus implying that
thermal conduction could be important. In \S3, we extend the original
1-temperature self-similar solutions of Narayan \& Yi (1994) with a
saturated form of thermal conduction and study the effects of
conduction on the flow structure and properties. In \S4, we comment on
some of the limitations of our work, on possibilities for future work
and on potential additional consequences of thermal conduction for hot
accretion.

\section{Observational Constraints and Collisionality}

{\it Chandra} observations provide tight constraints on the density
and temperature of gas at or near the Bondi capture radius in Sgr A*
and several other nearby galactic nuclei. These observational
constraints have been used before to estimate the rate at which the
gas is captured by the black hole in these systems, following Bondi
theory (e.g. Loewenstein et al. 2001; Di Matteo et al. 2003; Narayan
2003). Here, we use these same constraints (taken from Loewenstein et
al. 2001; Baganoff et al. 2003; Di Matteo et al. 2003; Ho, Terashima
\& Ulvestad 2003) to calculate mean free paths for the observed
gas.\footnote{It should be noted that these constraints are not all
equally good, in the sense that {\it Chandra} has probably resolved
the gas capture radius in Sgr A* and M87 but not in the other nuclei.}
Galactic nuclei and the corresponding gas properties on $1''$ scales
are listed in Table~\ref{tab:one}: $n_{1''}$ is the gas number
density, $T_{1''}$ is the temperature, $R_1$ is the $1''$
size-equivalent at the nucleus distance, $l_1$ is the mean free path
for the observed $1''$ conditions and $R_{\rm cap}$ is the capture
radius, inferred from the gas temperature and the black hole mass in
each nucleus. We have used Spitzer's (1962) expressions for collision
times, a standard expression for the thermal speed and a Coulomb
logarithm $\ln \Lambda \simeq 20$ to recover the simple mean free path
scaling $l \simeq 10^4 (T^2/n)$~cm of Cowie \& McKee (1997; valid for
both electrons and protons in a 1-temperature gas). We have used this
relation to calculate $l_1$ values in Table~\ref{tab:one}.

The inferred mean free paths are in the few hundredths to few tenths
of the observed $1''$ scales ($l_1/R_1$ values in
Table~\ref{tab:one}). Assuming no change in the gas properties down to
the inferred capture radius, values of $l_1/R_{\rm cap} \gsim 0.1$ are
deduced in four of the six nuclei. One could also reformulate these
scalings by stating that the $1''$ mean free paths are systematically
$\gsim 10^{4-5} R_{\rm s}$, where $R_{\rm s}$ is the Schwarzschild
radius (the typical lengthscale associated with the accreting black
holes). These numbers strongly suggest that accretion will proceed
under weakly-collisional conditions in these systems.

The weakly-collisional nature of ADAFs has been noted before
(e.g. Mahadevan \& Quataert 1997), in the sense of collision times
longer than the gas inflow time, but these are model-dependent
statements. Direct observational constraints on the gas properties
near the capture radius favor a weakly-collisional regime of hot
accretion, more or less independently of the exact hot flow
structure. Whether the gas adopts an ADAF, CDAF or ADIOS type
configuration once it crosses the capture radius, it is expected to
become even more weakly-collisional as it approaches the black hole,
since the relative mean free path scales as $l/R \propto R^{-3/2-p}$
in these flows with a virial temperature profile and a density profile
$\rho \propto R^{-3/2+p}$. It should be noted, however, that tangled
magnetic fields provide a way of limiting the weak collisionality of
the flow, even though their efficiency at doing so is not well
known. We go back this important issue in \S4.

\begin{deluxetable}{lccccc} \tablecolumns{6}
\tablewidth{0pt} \tablecaption{\label{tab:one} Observational Constraints on Collisionality} \tablehead{\colhead{Nucleus} &
\colhead{$n_{1''}$} & \colhead{$T_{1''}$} &
\colhead{$R_{1}$} & \colhead{$l_1/R_1$} & \colhead{$l_1/R_{\rm cap}$} \\
 & (cm$^{-3}$) & ($10^7$~K) & (cm)& &} \startdata
Sgr A*    & $100$ &  $2.3$  & $1.3 \times 10^{17}$ & $0.4$ & $0.4$\\
NGC 1399  & $0.3$ &  $0.9$  & $3.1 \times 10^{20}$ & $0.009$ & $0.02$\\
NGC 4472  & $0.2$ &  $0.9$  & $2.5 \times 10^{20}$ & $0.016$ & $0.07$\\
NGC 4636  & $0.07$&  $0.7$  & $2.2 \times 10^{20}$ & $0.032$ & $0.6$\\
M87       & $0.17$&  $0.9$  & $2.7 \times 10^{20}$ & $0.018$ & $0.02$\\
M32       & $0.07$&  $0.4$  & $1.2 \times 10^{19}$ & $0.2$   & $1.3$
\enddata
%\tablecomments{}
\end{deluxetable}

\section{Self-Similar Solution with Saturated Conduction}

The standard Spitzer formula for thermal conduction applies only to
gas well into the collisional (``fluid'') regime, with a mean free
path $l \ll L$, for any relevant flow scale $L$. Anticipating
temperature variations on local scales $\sim R$ from self-similarity,
in the weakly-collisional regime of interest here, with $l \sim R$ or
even $l \gg R$, a saturated (or equiv.  ``flux-limited'') thermal
conduction formalism must be adopted if one is to avoid unphysically
large heat fluxes. We adopt the formulation of Cowie and McKee (1977)
and write the saturated conduction flux as $F_s = 5 \Phi_s \rho
c_s^3$, where $\phi_s$ is the saturation constant (presumably $\lsim
1$), $\rho$ is the gas mass density and $c_s$ is its sound speed. With
this prescription, the largest achievable flux is approximated as the
product of the thermal energy density in electrons times their
characteristic thermal speed (assuming a thermal distribution and
equal electron and ion temperatures; see Cowie and McKee 1977 for
details). Because it is a saturated flux, it no longer explicitly
depends on the magnitude of the temperature gradient but only on the
direction of this gradient. Heat will flow outward in a hot accretion
flow with a near-virial temperature profile, hence the positive sign
adopted for $F_s$.

With this simple formulation for conduction, the steady-state
1-temperature self-similar ADAF solution of Narayan \& Yi (1994) can
be generalized to include the divergence of the saturated conduction
flux in the entropy-energy equation.\footnote{Honma (1996) and Manmoto
et al. (2000) have investigated the role of ``turbulent'' heat
transport in ADAF-like flows. By relying on a saturated form of
``microscopic'' thermal conduction which is physically well-motivated,
our analysis differs substantially from theirs. Saturation is key to a
self-similar scaling, which would not obtain for a standard Spitzer
conduction law.}  Adopting the same geometry and notation as Narayan
\& Yi (1994), we consider a ``visco-turbulent'',
differentially-rotating hot flow which satisfies the following
height-integrated equations for the conservation of momentum and
energy
\begin{eqnarray}
v \frac{dv}{dR} &=& R (\Omega^2-\Omega_K^2)-\frac{1}{\rho}\frac{d}{dR}(\rho c_s^2),\\
v \frac{d(\Omega R^2)}{dR} &=& \frac{1}{\rho R H} \frac{d}{dR} \left( \frac{\alpha \rho c_s^2 R^3 H}{\Omega_K} \frac{d \Omega}{dR}\right),\\
2 H \rho v T \frac{ds}{dR} & = & f \frac{2 \alpha \rho c_s^2 R^2 H}{\Omega_K} \left( \frac{d \Omega}{dR}\right)^2 - \frac{2H}{R^2} \frac{d}{dR} (R^2 F_s),\label{eq:ent}
\end{eqnarray}
supplemented by the continuity equation, which is rewritten as $\dot
M= -4 \pi R H v \rho$. The last term in Eq~(\ref{eq:ent}) is the
additional divergence of the saturated conduction flux.  In these
equations, $R$ is the cylindrical radius, $\Omega_K$ is the Keplerian
angular velocity around the central gravity-dominating mass, $v$ is
the gas radial (in)flow speed, $\Omega$ is its angular velocity,
$\rho$ is the mass density, $c_s$ is the isothermal sound speed, $s$
is the gas specific entropy, $T$ is its temperature, $H= R c_s/v_K$ is
the flow vertical height and $f \leq 1$ is the advection parameter
($f=1$ for negligible gas cooling). A Shakura-Sunyaev prescription has
been used to capture the ``visco-turbulent'' nature of the flow, with
an equivalent kinematic viscosity coefficient $\nu =\alpha
c_s^2/\Omega_K$, where $\alpha$ is the traditional viscosity
parameter.

We look for solutions to these equations of the form
\begin{equation}
\rho = \rho_0 R^{-3/2},~v=v_0 R^{-1/2},~\Omega=\Omega_0 R^{-3/2},~c_s^2=c_{s0}^2 R^{-1},  
\end{equation}
and use the notation $v_K=v_{K0} R^{-1/2}=R \Omega_K=R \Omega_{K0}
R^{-3/2}$.  We find that the three unknowns $v_0$, $\Omega_0$ and
$c_{s0}$ satisfy the relations
\begin{eqnarray}
\frac{1}{2}v_0^2 &-& \Omega_{K0}^2 + \Omega_0^2 + \frac{5}{2}c_{s0}^2=0,\label{eq:bern}\\
v_0 &=& -\frac{3 \alpha}{2} \frac{c_{s0}^2}{v_{K0}},\\
\frac{9 \alpha^2}{8} c_{s0}^4 &+& \left( \frac{5}{2}+\frac{5/3-\gamma}{f(\gamma-1)} \right)c_{s0}^2 v_{K0}^2 -\frac{20 \phi_s}{9 \alpha f}c_{s0} v_{K0}^3 - v_{K0}^4 =0,\label{eq:enrel}
\end{eqnarray}
while the density scale, $\rho_0$, is fixed by the accretion rate
$\dot M$.

Contrary to the original ADAF solution, the energy relation
(Eq.~[\ref{eq:enrel}]) is not a quartic in $c_{s0}/v_{K0}$ because of
the extra saturated conduction term. We solve this fourth order
polynomial with a standard numerical technique (Press et al. 1992) and
discard the two imaginary roots as well as the real root with negative
$c_{s0}$. While in the original ADAF solution the two real roots for
$c_{s0}$ were degenerate (in the sense that only $c_s^2$ mattered),
this degeneracy is lifted up with saturated thermal conduction. The
root with negative $c_{s0}$ is physically unacceptable because it
leads to a conduction flux going up the temperature gradient. Note
that we have focused on accretion solutions ($v_0 < 0$) with $1 <
\gamma < 5/3$ in our search, but by analogy with Narayan \& Yi (1994),
one also expect rotating wind solutions ($v_0 > 0$) with saturated
thermal conduction to exist for values of $\gamma > 5/3$.

\begin{figure}
\centering\mbox{\includegraphics[width=8cm]{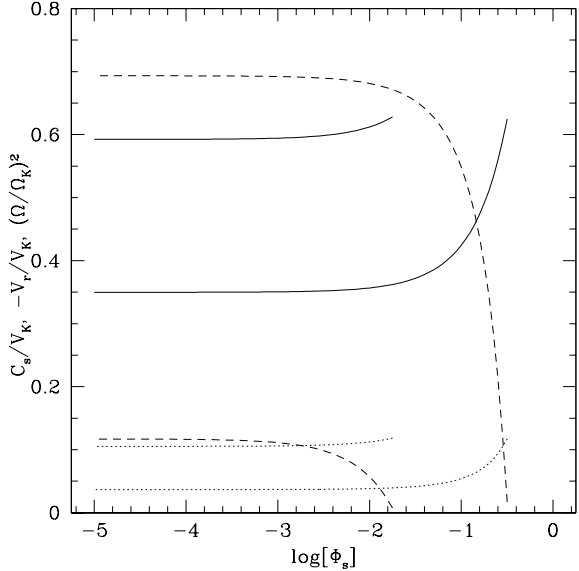}}
\caption[]{\label{fig:one} Changes in the sound speed ($c_s/v_K$,
solid lines), angular rotation velocity ($\Omega^2 / \Omega_K^2$,
dashed lines) and radial inflow speed ($- v_r/v_K$, dotted lines) for
two distinct solutions, as a function of the saturation constant,
$\Phi_s$. At low $\Phi_s$ values, these solutions match a regular
ADAF. The solution extending up to $\log \Phi_s \simeq -0.5$ has a gas
adiabatic index $\gamma = 1.1$, while the solution extending up to
$\log \Phi_s \simeq -1.9$ has $\gamma =1.5$.  Both solutions have
$\alpha=0.2$ (viscosity parameter) and $f=1$ (advection parameter).}
\end{figure}

The main features of the new self-similar solutions with saturated
conduction are illustrated in Fig.~\ref{fig:one}, as a function of the
saturation constant, $\Phi_s$. We are showing results for two distinct
solutions here, with $\gamma=1.5$ and $\gamma=1.1$, and have fixed
$f=1$ and $\alpha =0.2$ in both cases for definiteness. Standard ADAF
solutions are recovered at small $\Phi_s$ values. As $\Phi_s$
approaches unity, however, the solutions deviate substantially from
the standard ADAF. The sound speed, $c_s$, and the radial inflow
speed, $-v_r$, both increase with the magnitude of conduction, while
the squared angular velocity, $\Omega^2$, decreases. Solutions still
exist when $\Omega^2$ becomes negative but they are obviously
physically unacceptable. The solution with $\gamma=1.5$ reaches the
non-rotating limit at $\log \Phi_s \simeq -1.9$, while the solution
with $\gamma =1.1$ reaches this limit for $\log \Phi_s \simeq
-0.5$. Both solutions have a sub-virial temperature ($c_s< v_k$) and a
subsonic radial inflow speed ($-v_r< c_s$), which appear to be rather
general properties. Supersonic solutions may exist for large $\alpha$
values when $\gamma \to 5/3$.

As the level of saturated conduction is increased, more and more heat
flows from the hotter, inner regions, resulting in a local increase of
the gas temperature relative to the original ADAF
solution. Simultaneously, the gas adjusts its angular velocity (which
reduces the level of viscous dissipation) and increases its inflow
speed to conserve its momentum balance (increasing at the same time
the level of advection). One can show from the energy equation that
$\Omega^2 \propto (1-|q_{\rm cond}/q_{\rm adv}|)$, where $q_{\rm
cond}$ and $q_{\rm adv}$ are the magnitudes of the conduction and
advection terms: solutions cease to exist once advection becomes
unable to balance the heating due to saturated conduction.  Solutions
with $\Omega^2 = 0$ can be seen as Bondi-like with saturated
conduction.

We have found that variations in the advection parameter, $f$, have
little effect on the solutions, as long as $f \lsim 1$. The breakdown
of the solutions (when $\Omega^2 \to 0$) occurs at lower values of the
saturation constant $\Phi_s$ for smaller $\alpha$ and larger $\gamma$
values. While this breakdown may seem problematic, because it occurs
for smallish values of $\Phi_s$ when $\gamma \lsim 5/3$
(Fig.~\ref{fig:one}), it may well turn out to be a simple pathological
feature of the 1D height-integrated equations we solved. By analogy
with the results of Narayan \& Yi (1995a) on convection in 2D
self-similar ADAF solutions, we might expect conduction to dominate
over advection only in the polar regions of the flow, in which case it
may be possible to extend the solutions to larger $\Phi_s$ values (and
perhaps launch outflows).

\begin{figure}
\centering\mbox{\includegraphics[width=8cm]{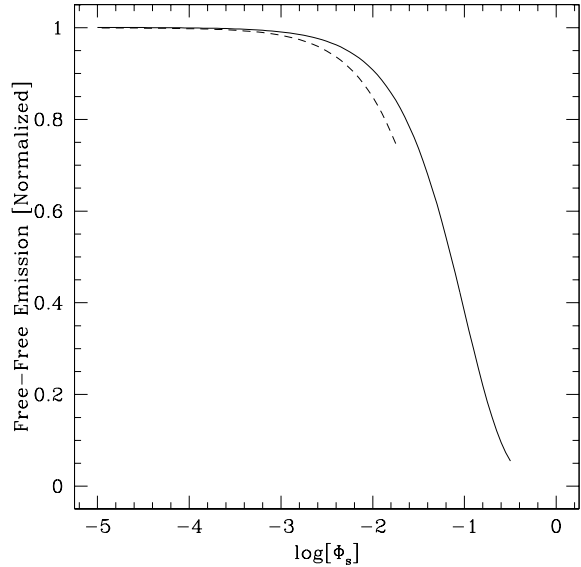}}
\caption[]{\label{fig:two} Drop in the level of free-free emission,
relative to the standard ADAF value, as a function of the saturation
constant, $\Phi_s$, for the same two solutions as in
Fig.~\ref{fig:one} and a fixed accretion rate, $\dot M$.}
\end{figure}

From the solution scalings, it is possible to calculate the modified
level of emission expected relative to a standard ADAF. Assuming that
free-free emission is the dominant mechanism, it scales as $\rho^2
T^{1/2} \propto (c_{s0} v_0^2)^{-1}$ for a given $\dot M$. Changes in
this quantity as a function of the saturation constant, $\Phi_s$, for
the two same specific solutions of Fig.~\ref{fig:one}, are shown in
Fig.~\ref{fig:two} (normalized to the low-conduction$=$ADAF
values). The emission is effectively reduced, potentially by a large
factor (specifically by $\sim 20$ in our $\gamma =1.1$ model). Even
though conduction heats up the gas locally, the reduced density
resulting from the larger inflow speed dominates, leading to a net
decrease in the expected level of free-free emission. Clearly, by
reducing the radiative efficiency at a given $\dot M$, saturated
conduction could help explain the phenomenology of dim accreting black
holes. However, work beyond these idealized self-similar solutions
will be required before we can elucidate the role of conduction in hot
accretion flows.

\section{Discussion and Conclusion}

The possibility that tangled magnetic fields strongly limit the
efficiency of thermal conduction in a hot flow remains a major
theoretical uncertainty. In a sense, we have accounted for this
possibility by allowing the value of the saturation constant,
$\Phi_s$, to vary in our solutions. In recent years, a similar issue
has been discussed in the cooling flow context: several studies have
concluded that magnetic fields would probably not strongly limit
thermal conduction (e.g. Narayan \& Medvedev 2001; Gruzinov 2002; see
also Chandran \& Cowley 1998; Chandran et al. 1999), but it remains
somewhat of an open issue. Explicit numerical simulations with
anisotropic conduction along fields in an MHD turbulent medium could
greatly help in settling this issue (see, e.g., Cho \& Lazarian 2004).

An important limitation of the self-similar solutions with saturated
conduction is their 1-temperature structure. Narayan \& Yi (1995b)
have shown how the 2-temperature property of the hot flow is crucial
to obtain self-consistent solutions with realistic cooling
properties. The decoupling of the electron and ion temperatures would
modify the role of saturated conduction in the inner regions of the
hot flow, an effect which is not captured by our solutions. In this
situation, the ions may be insulated from electron conduction. In
addition, self-similarity is broken in the 2-temperature regime and
the electron temperature profile flattens to a sub-virial slope
(e.g. Narayan, Mahadevan \& Quataert 1998). While this suggests a
reduced role for conduction, the break-down of self-similarity also
implies that the heating of the outer flow regions must be done at the
expense of the hottest inner regions. Given the strong dependence on
temperature of cyclo-synchrotron emission, even a moderate drop in
temperature in the hottest regions could lead to a significant
reduction in the radio emission properties of the flow. More detailed
models are needed to address these issues and determine quantitatively
by how much conduction can contributes to the low-radiative efficiency
of dim accreting black holes.

In the last few years, much emphasis has been put on reducing the rate
at which gas in a hot flow is accreted by the black hole, either by
invoking a positive Bernoulli constant to justify the existence of
powerful outflows (Blandford \& Begelman 1999) or by invoking a net
reduction in the accretion rate due to convection (Narayan et
al. 2000; Quataert \& Gruzinov 2000). The presence of strong
conduction in the flow may pose a challenge to both of these
scenarios: if conduction is important on all scales of interest in the
hot flow (because of the large mean free paths), the Bernoulli
constant\footnote{Satisfying the same Eq~(\ref{eq:bern}) as ADAFs
(Narayan \& Yi 1994), solutions with saturated conduction have
formally larger positive Bernoulli constant because of an increased
temperature and reduced angular velocity.} becomes irrelevant (as it
is a property of adiabatic flows) and convection may be suppressed
when any displaced fluid element efficiently reaches thermal
equilibrium with its environment through conduction (but see Balbus
2001). Obviously, this is not to say that outflows are not expected
from hot accretion onto black holes, since different configurations
than the ones described by idealized steady-state analytical solutions
are certainly possible (as shown explicitly by global turbulent MHD
simulations).

If conduction is indeed important for hot accretion onto black holes,
it could also have other interesting consequences besides the ones
illustrated in our specific solutions. For instance, if the hot flow
inside of the capture radius around a black hole is able to heat up
the ambient gas outside of that capture radius, the Bondi rate of gas
capture may be effectively reduced. While Gruzinov (1998) invoked
``turbulent'' heat flows to achieve this, the weakly-collisional
conditions discussed in \S2 for several galactic nuclei suggest that
this could be done by ``microscopic'' conduction itself. A reduced
rate of gas capture below the Bondi value could go a long way in
explaining the low luminosity of these accreting black holes, but here
again detailed models are required to address this issue
quantitatively.  Conduction could also have important consequences for
the dynamical structure of hot flows, as argued recently by Balbus
(2001) in a stability analysis in the presence of anisotropic
conduction. In addition, in a weakly-collisional hot flow, it is in
principle possible for the ``microscopic'' viscous stress tensor to
contribute significantly to the overall flow dynamics, along with the
turbulent component usually emphasized in accretion theory
(e.g. Quataert et al. 2002).
 
Finally, conduction may offer a solution to the disk evaporation
problem. Around both stellar-mass and supermassive black holes, there
is evidence for inner truncation of thin accretion disks (e.g. Esin et
al. 2001; Quataert et al. 1999). The process which ``evaporates'' the
inner disks into a hot flow is not well understood, however. A simple
calculation shows that, independently of the mass of and distance from
the black hole, the saturated conduction flux ($\sim \rho c_s^3$)
available to heat up a thin disk embedded in a hot flow (using ADAF
$\rho$ and $c_s$ scalings for simplicity; Narayan et al. 1998) is
within $\sim \alpha^{-1}$ of the traditional viscous dissipation rate
per unit area of the disk (for a same $\dot M$ in both components).
This suggests that saturated conduction is energetically capable of
contributing to the evaporation of thin disks in both classes of
systems.

\vspace{0.3cm}

The author is grateful to S. Balbus, W. Dorland and E.Quataert for
useful discussions.  This research was supported in part by the
National Science Foundation under Grant No. PHY99-07949 (at KITP).

\end{document}